%% file: main.tex
\documentclass[twocolumn]{aastex631}

\usepackage{float}
\shorttitle{Towards a data-driven model of the sky from low Earth orbit}
\shortauthors{Caddy et al.}

\graphicspath{{./}{figures/}}

\begin{document}

\title{Towards a data-driven model of the sky from low Earth orbit as observed by the Hubble Space Telescope}

\correspondingauthor{Sarah E. Caddy}
\email{sarah.caddy@mq.edu.au}

\author[0000-0001-6990-7792]{Sarah E. Caddy}
\affil{School of Mathematical and Physical Sciences, Macquarie University, Sydney, NSW 2109, Australia}
\affil{Research Centre in Astronomy, Astrophysics \& Astrophotonics, Macquarie University, Sydney, NSW 2109, Australia}
\author[0000-0001-5185-9876]{Lee R. Spitler}
\affiliation{School of Mathematical and Physical Sciences, Macquarie University, Sydney, NSW 2109, Australia}
\affiliation{Research Centre in Astronomy, Astrophysics \& Astrophotonics, Macquarie University, Sydney, NSW 2109, Australia}
\affiliation{Australian Astronomical Optics, Faculty of Science and Engineering, Macquarie University, Macquarie Park, NSW 2113, Australia}
\author[0000-0002-0742-379X]{Simon C. Ellis}
\affiliation{Research Centre in Astronomy, Astrophysics \& Astrophotonics, Macquarie University, Sydney, NSW 2109, Australia}
\affiliation{Australian Astronomical Optics, Faculty of Science and Engineering, Macquarie University, Macquarie Park, NSW 2113, Australia}

%



\begin{abstract}

The sky observed by space telescopes in Low Earth Orbit (LEO) can be dominated by stray light from multiple sources including the Earth, Sun and Moon. This stray light presents a significant challenge to missions that aim to make a secure measurement of the Extragalactic Background Light (EBL). In this work we quantify the impact of stray light on sky observations made by the Hubble Space Telescope (HST) Advanced Camera for Surveys. By selecting on orbital parameters we successfully isolate images with sky that contain minimal and high levels of Earthshine. In addition, we find weather observations from CERES satellites correlates with the observed HST sky surface brightness indicating the value of incorporating such data to characterise the sky. Finally we present a machine learning model of the sky trained on the data used in this work to predict the total observed sky surface brightness. We demonstrate that our initial model is able to predict the total sky brightness under a range of conditions to within $3.9\%$ of the true measured sky. Moreover, we find that the model matches the stray light-free observations better than current physical Zodiacal light models.

\end{abstract}


\input{01_intro}
\input{02_method}

\input{03_results}

\input{04_discussion}

\input{05_conclusion}

\begin{acknowledgments}
We would like to thank Prof. Rogier Windhorst, Dr Timothy Carleton, Rosalia O'Brien and the entire SKYSURF collaboration for their continued support of this work. We would also like to thank the Huntsman Telescope team for their support and guidance. We would like to thank the staff at the HLA for their help sourcing the HST ACS data used in this work, and the staff at NORAD for compiling HST ephemris data for this work on special request. We would also like to thank Dr Grant Matthews from the CERES collaboration for his guidance using CERES data, and for providing the CERES throughput tables for this work. Finally, We would like to thank Arvind Hughes for his helpful comments which led to the use of XGBoost for this work. This work is made possible by observations made with the NASA/ESA Hubble Space Telescope, and obtained from the Hubble Legacy Archive, which is a collaboration between the Space Telescope Science Institute (STScI/NASA), the Space Telescope European Coordinating Facility (ST-ECF/ESAC/ESA) and the Canadian Astronomy Data Centre (CADC/NRC/CSA). SC and LS acknowledge support from an Australian Research Council Discovery Project grant (DP190102448).
\end{acknowledgments}

\software{Skypy (Caddy et al. 2022), 
Gunagala (Horton et al. 2021),
XGBoost (The XGBoost Contributors, 2021),
Astropy (The Astropy Collaboration 2013, 2018),
PySynphot (The Space Telescope Science Institude, 2021),
PyEphem (Rhodes 2020),
}

\clearpage
\appendix

\begin{table}[!h]\label{sec:fits_header_params}
\caption{Description of StarView key words and HST FITS header keywords used in this work}
\begin{tabular}{@{\vrule height 14pt depth 6pt width 0pt}llllp{3in}}
Data Start Time &=&  The start time and date of the exposure \\
Data End Time   &=&   The end time and date of the exposure \\
Target Name      &=&   Proposer’s target name \\
EXPTIME          &=& Exposure time (s) \\
FILTER1        &=& Element selected from filter wheel 1   \\
FILTER2         &=& Element selected from filter wheel 2 \\
GLAT\_REF       &=& Galactic latitude of the target (deg) (J2000)\\
GLON\_REF       &=& Galactic longitude of the target (deg) (J2000)\\
PHOTMODE        &=& Combination of \textless{}INSTRUMENT\textgreater{}+\textless{}CAMERA\textgreater{}+\textless{}FILTER\textgreater{}\\ 
PHOTPLAM        &=& Pivot wavelength of the photmode ($\AA$) \\
PHOTFLAM        &=& Inverse sensitivity (erg/$cm^{2}$/$\AA$/DN) \\
PHOTZPT         &=& ST magnitude system zero point   \\
PHOTBW          &=& Root Mean Square bandwidth of the photmode ($\AA$)   \\
MDRIZSKY        &=& Astrodrizzle median sigma clipped sky surface brightness (DN/pix)    \\ 
SUNANGLE        &=& Angle between Sun and V1 axis (deg) \\
SUN\_ALT        &=& Altitude of the Sun above earth’s limb (deg) \\
LOS\_MOON       &=& Angle between the Moon and V1 axis (deg)\\
ALTITUDE        &=& The average altitude of the telescope (km)  \\
\end{tabular}
\end{table}

\clearpage
\bibliography{refs}{}
\bibliographystyle{aasjournal}



\end{document}

%% file: 01_intro.tex
\section{INTRODUCTION}\label{the_start}
The total sky observed by a low Earth orbiting (LEO) space telescope can contain significant levels of``stray light'' which contains complex and time-varying sources like Earthshine, Sunshine and Moonshine as well as bright stars outside the optical field of view. In this work we explore whether we can characterise the observing conditions where stray light is most likely to impact observations from LEO, and predict the total intensity of the sky in order to support science goals to measure both diffuse and low surface brightness signatures that require precise sky photometry.
\\
\\
A number of LEO space telescopes (e.g., SPHEREx \citealt{Korngut2018}; SkyHopper\footnote{\url{https://skyhopper.research.unimelb.edu.au/}}, MESSIER \citealt{Lombardo2019}), sounding rocket missions (e.g., CIBER \citealt{Korngut2013}, CIBER2 \citealt{Korngut2021}), and large archival surveys from existing instruments (e.g., SKYSURF \citealt{Windhorst2022, Carleton2022}) must deal with sources of stray light in order to make a secure measurement of the extragalactic background light (EBL). The EBL consists of all optical/infrared sources over all cosmic time and is considered a valuable benchmark to assess competing theories of structure formation of the Universe. 
\\
\\
Attempts to make a secure measurement of the EBL have resulted in a significant unresolved discrepancy between direct measurements and estimates derived from integrated galaxy counts \citep[see][for a review]{Cooray2016}. If the EBL is derived from galaxies, these estimates should converge \citep{Driver2016}. However, this is not yet found to be the case \citep{Mattila1975,Spinrad1978,Boughn1986,Mattila1990,Bernstein,Matsuoka2011,Korngut2013, Driver2016,Mattila2017,Mattila2019,Korngut2021, Lauer2022}. Current best estimates from direct measurements are up to a factor of five times larger then estimates from galaxy counts \citep{Driver2016}. Indirect measurements made by studying the signature of the EBL on the of spectra of bright gamma ray sources also suffer from systematic errors \citep{Korngut2021}. However it provides another independent detection of a lower limit that leaves little room for the presence of unresolved sources \citep{Desai2019}. In contrast, measurements made by \cite{Lauer2022} with New Horizons LORRI images in the outer solar system reported a potential detection of excess light that was not explained by resolved galaxies \citep{Korngut2021}. 
This discrepancy may indicate the presence of a truly diffuse component of the EBL undetected by galaxy counts. It could also indicate an incompleteness in the model of sky brightness from LEO that may lead to overestimated sky subtractions in estimates derived from direct EBL observations by space telescopes \citep[see][for examples]{Windhorst2022}.
\\
\\
Direct measurement of the EBL from all {\it unresolved} optical/infrared sources \citep{Matsuoka2011,Driver2016,Zemcov2017,Mattila2019}, has been challenging to isolate from the numerous other components (including stray light) that make up the total observed sky. The observed sky from UV to near-IR can be broken into several components. Each component originates from a different source, but once combined they are difficult to separate because the sources do not have strongly defining spectral features that can be used to distinguish one from another. The components of the diffuse sky from LEO include 4 main contributions:
\begin{itemize}
    \item Zodiacal Light, composed of both scattered and thermally radiated Sun light from interplanetary dust particles distributed non-uniformly in the Solar System \citep{Kelsall1998}. 
    \item The diffuse galactic light, composed of scattered light from dust and gas in the Milky Way's interstellar medium \citep{Leinert1998}. 
    \item The EBL, composed of all radiating and re-emitting sources integrated over all space, time and wavelengths \citep{Driver2016}. The EBL can be described by two components: the diffuse EBL comprised of unresolved sources, and the discrete EBL consisting of resolvable sources. 
    \item Stray light, which here is defined as any light scattered or emitted from outside the field of view of the instrument. It includes Earthshine, Moonshine, or Sunlight. It also includes other bright astronomical sources like stars \citep[e.g.][]{Borlaff2021}, which we do not address in this work specifically. 
\end{itemize}
The two brightest and most variable components of the sky are the Zodiacal Light and stray light. Measuring these components poses the most challenging barrier in securing a measurement of the EBL from LEO \citep[e.g. ][]{Korngut2021}. 
\\
\\
Efforts to constrain Zodiacal Light have resulted in time-varying models that describe the Zodiacal Light surface brightness over the entire sky using data from COBE/DIRBE \citep{Kelsall1998}, Spitzer \citep{Krick}, and/or SMEI \citep{Buffington2016}. The two most commonly used models are the Wright model \citep{Wright2001}, and the Kelsall model \citep{Korngut2021}. A limitation of these models is their reliance on poorly-constrained models of the interplanetary dust distribution within the Solar System \citep{Jorgensen2021}. Indeed, the Kelsall and Wright models disagree by as much as the total intensity of the expected EBL itself \citep{Korngut2021, Driver2016}. This systematic uncertainty makes it challenging to distinguish between the EBL and the Zodiacal Light sky.
\\
\\
Stray light from Earthshine and Sunshine can be an important component of the total observed sky for space telescopes in LEO observing at optical wavelengths, because the sources are so luminous. At $\sim0.3$ - $2 \mu$m Earthshine is dominated by reflected Sunlight and is characterized by a scattered Solar spectrum with the addition of absorption features from the atmosphere, clouds, forests, oceans and deserts \citep{Giavalisco2002, Doelling2013}. As a result, the stochastic nature of the weather combined with time-varying surface features beneath a space telescope makes Earthshine stray light difficult to model accurately.
\\
\\
The easiest solution to mitigate the impact of stray light from Sunshine and Earthshine for observers is to simply avoid it by restricting the range of the telescope's pointing with respect to potential stray light sources \citep{Shaw1998, Giavalisco2002, Korngut2018}. This comes at a cost to the productivity of the instrument, reducing the visit time for particular targets. For this reason pointing restrictions are usually not designed to completely avoid Earthshine or Sunshine stray light, only minimise it. Indeed, using the dataset in the present study, we estimated that potentially more than half of the exposures in the Hubble Legacy Archive (HLA) may be contaminated by some sort of stray light based on the Sun Altitude at which the exposure was taken.
\\
\\
Various studies have characterised stray light from space telescopes. For example, \cite{Murthy2014} use archival imaging products from the GALEX space telescope to find two components in the UV sky: one that is time-variable symmetrical in intensity about the local midnight of the orbiting observatory, and one that is proportional to the angle between the Sun and the observed field. \citet{Borlaff2021} used Hubble Space Telescope (HST) stray light information in the Hubble Ultra Deep Field (HUDF) to better inform sky estimates for future missions such as Euclid and JWST. Prior work to understand how HST orbital parameters and telescope attitude impact the presence of stray light has led to rough estimates of the intensity of Earthshine stray light contributions \citep{Shaw1998, Giavalisco2002, Biretta2003, Baggett2012, Brammer2014}. Some of this work informs the 3 options currently available in the HST Exposure Time Calculator for Earthshine contribution (average, high, or extremely high) with an important caveat that these often do not reflect true conditions during operations \citep{Giavalisco2002}.
\\
\\
The impact of Earthshine on space based telescopes in LEO is nicely illustrated by the work of \cite{Luger2019}. They detect Earthshine stray light variations in Transiting Exoplanet Survey Satellite \citep[TESS; ][]{Borucki2010} data and use it to map the continents on the surface of the Earth. In doing so they demonstrate that TESS is an ``effective addition to NASA’s fleet of weather satellites''. In this work we explore this idea further and aim to answer the question: can NASA's fleet of weather satellites be used to inform the impact of Earthshine on LEO space telescopes? 
\\
\\
In this work we create a data-driven model to assess and predict the impact of stray light. The analysis is used inform when HST data is likely to have minimal to no stray light from the Earth, Sun and Moon. We also illustrate the promising features of having a data-driven model to support EBL science goals. A future version of the model might be used to improve exposure time calculations by using a highly-tuned prediction for the sky surface brightness for a given field observed at a particular time with HST or future missions.
\\
\\
In detail, here we extend the work of \citet{Caddy2019} and construct empirically-derived model of the sky brightness observed by HST. We leverage the flexibility and accuracy of the machine learning algorithm XGBoost \citep{Chen2016} and the extensive data of the Hubble Legacy Archive (HLA) composed of hundreds of thousands of exposures - spanning decades - in multiple filters, to create a useful tool that aims to predict stray light from LEO. We describe the results of the constructed models using calculated and collated orbital parameters of HST, the median clipped sky in a sample over 34,000 Advanced Camera for Surveys Advanced Camera for Surveys (ACS) \citep{Sirianni2005} images, and the Earthshine below the space telescope derived from simultaneous satellite imagery from the CERES missions. We explore the impact of each pointing parameter on the contribution of local stray light to the total sky, and demonstrate the benefits of constructing an empirically generated sky model that incorporates all foreground stray light sources, as opposed to identifying and modeling each component of the sky separately. Finally we demonstrate the use and functionality of the model utilising GOODS North, and describe future work made possible by these results.

%% file: 02_method.tex
\section{Method}\label{sec:method}
In this section we describe the observational data and derived quantities used in this work. We also describe the Earthshine observations made by orbiting weather satellites, and how these data are translated to fluxes beneath HST. The calibration of the HST data, HST data quality control and the construction of a geometric model describing HST's pointing relative to the Earth are presented, and we describe the XGBoost machine learning model used in this work to predict the total intensity of the sky.
\begin{figure*}[t!!!!]
    \centering
    \includegraphics[width=18cm]{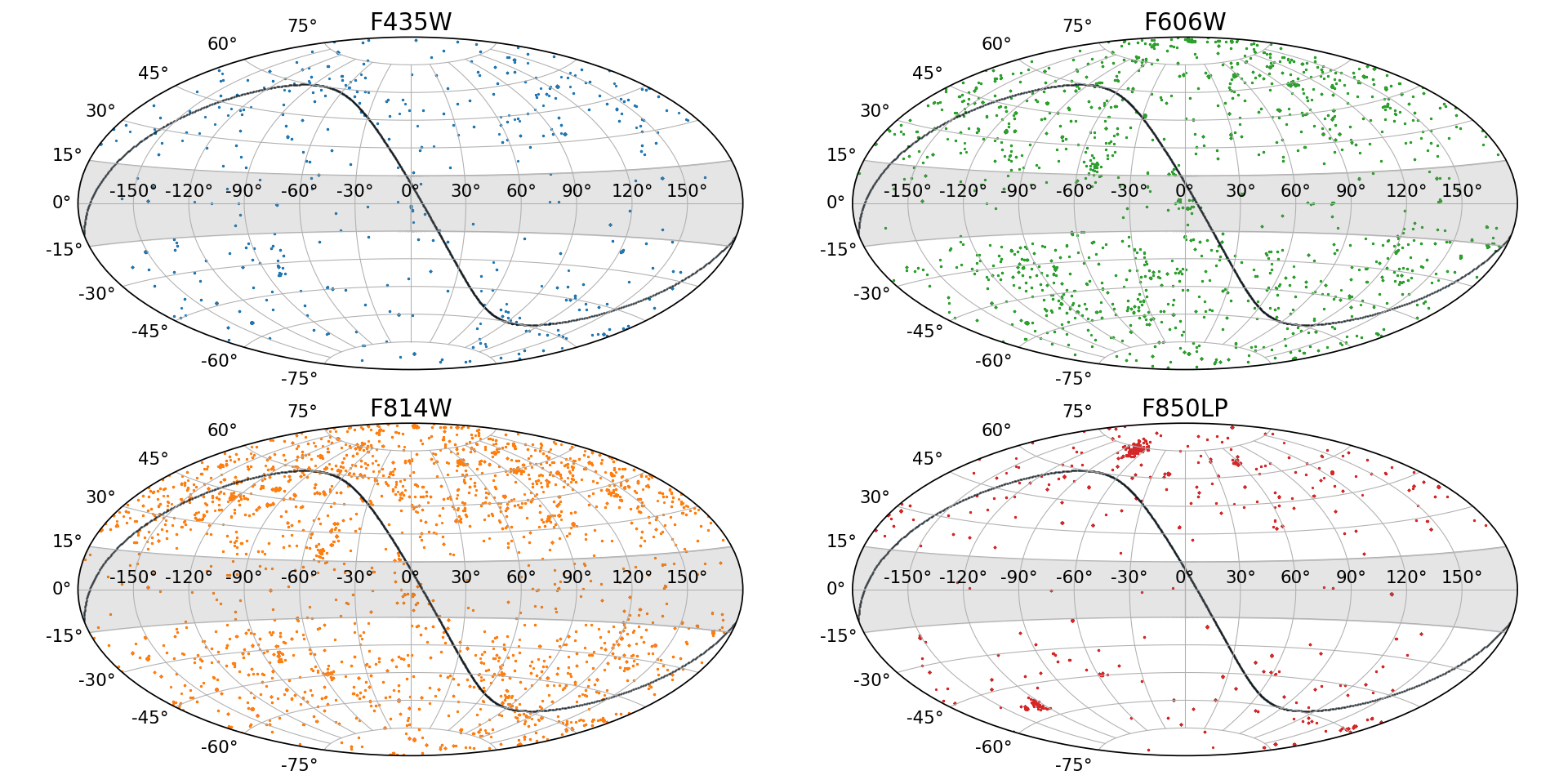}
    \caption{The location of fields used in this study in 4 selected filters. Field location is plotted in galactic coordinates, with the galactic plane from $\pm 15^{\circ}$ shown in shaded grey. The ecliptic plane is plotted in black. ACS F435W, F606W, and F814W are chosen due to the quantity of data available with existing archival sky surface brightness estimates over the entire sky. ACS F850LP is chosen because of data in the GOODS North fields which are known to have high levels of Earthshine contamination.}
    \label{fig:field_locations}
\end{figure*}
\subsection{Astrodrizzle sky estimates}\label{drizz}
\begin{figure*}[ht!!!!!!]
    \centering
    \includegraphics[width=18cm]{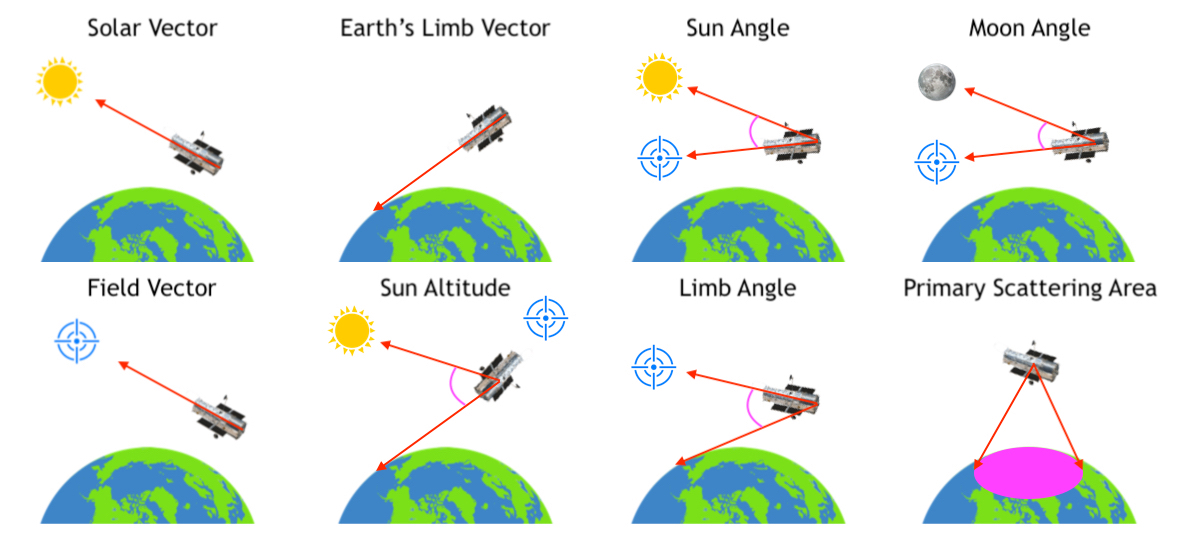}
    \caption{In this study four key angles are used to describe the relationship between the orientation of the telescope and potential sources of stray light. The four angles and the vectors that define them are illustrated in this figure. The primary scattering area is from a circular region defined by the visible horizon from a given orbital altitude. For this work angles are defined from the midpoint of a given exposure with respect to the surface of Earth, the Moon or the Sun.}
    \label{fig:angles_and_vectors}
\end{figure*}

In this study we use data derived from $\sim$ 34,300 individual exposures taken by the ACS on HST from 2002 to 2020. We use only ACS data because sky estimates are readily available in the (now deprecated) StarView engineering database \citep{Holl1992}. Four filters are used for this study (F453W, F606W, F814W and F850LP) due to the large amount of available data to sample the optical stray light and Zodiacal Light. The F850LP GOODS North \citep{Dickinson2006} data is particularly useful because early observations are known to have high levels of stray light contamination \citep{Kawara2014}. The locations of the fields on sky are shown in galactic coordinates in \autoref{fig:field_locations} and, aside from poor sampling near the galactic plane, are relatively well distributed over time and space.
\\
\\
Attitude parameters define the orientation of the telescope's axes with respect to the Earth, Sun and Moon, and are used as indicators of stray light contamination. These variables are described in \autoref{fig:angles_and_vectors} and are retrieved for every available ACS exposure in standard FITS headers from the database StarView. The relevant FITS header keywords and Star View key words are summarised in \autoref{sec:fits_header_params}. 
\\
\\
The median sky surface brightness estimated for each exposure is taken from the FITS header keyword \textbf{MDRIZSKY}, which is computed by an automated sky subtraction routine in Astrodrizzle in STScI Drizpack software \citep{Hack2019}. These FITS header keywords were obtained from the StarView database. The \textbf{MDRIZSKY} is computed by sigma clipping pixels with outlying values. After each of the five sigma clipping iterations, the standard deviation of pixel values is computed, and pixels deviating from the mean value by more than $4 \sigma$ are rejected. The median value of non-rejected pixels is the adopted estimate of the sky level. The number of pixels used in the calculation of the median sky value is not retained or recorded during the Drizzling process, so we cannot easily estimate uncertainties for individual measurements. We will address this shortcoming in a future work.
The sky estimates generated by Astrodrizzle are given in units of un-calibrated instrumental flux $e^{-}/\text{pixel}$ in ACS images. For each observation, the zeropoint can be derived for the specific instrument configuration given photometric header keywords. PHOTFLAM ($erg/cm^2/\AA/e^{-}$) for each filter is used as the scaling factor necessary to transform an instrumental flux in units of $e^{-}/\text{pixel}$ to a physical flux density. The subtracted sky estimate can then be represented in terms of physical flux density in units of $erg/s/cm^2/\text{arcsec}^2/\AA$ by dividing by the squared pixel scale of the instrument and the exposure time: 
\begin{equation}
    F_{\lambda} = \frac{(\text{MDRIZSKY} \times \text{PHOTFLAM})}{\text{EXPTIME} \times 0.05''^{2}}
\end{equation}
We use the original pixel scale of ACS/WFC of $0.05''/\text{pixel}$ because drizzling images to a range of different pixel scales generally occurs after \textbf{MDRIZSKY} is calculated from the physical instrument pixel scale \citep{Hack2019}.

\subsection{Sky Estimates Data Quality Control} \label{data_quality_control}

Particularly crowded fields such as star clusters, planetary targets, and large foreground targets such as NGC objects that take up the entire field of view and can corrupt the automatically generated \textbf{MDRIZSKY} sky estimates. For this reason fields containing these targets were removed from the dataset by screening the StarView field name {\textbf{Target Name}}, and in some cases confirming this information with the proposal ID available on the HLA. Fields with exposure times of less than 500 seconds are also removed to reduce the impact of charge transfer efficiency losses for particularly faint sky observations. Exposure times are limited to less than 2000s to reduce the impact of longer duration exposures sampling both day and night sides of the orbit in one exposure. Duplicated data are removed with identical start times and keyword parameters. Calibration fields or unknown targets that are sometimes returned in error from the StarView database were also removed from the analysis.
\\
\\
Unfortunately due to unknown reasons, the information in FITS headers provided on StarView are not always correct or complete. As a result, other outliers that may impact the accuracy of the measured sky includes grism images, calibration images, and images with failed guide star acquisition. Every effort has been made to pick out these images by manual inspection of the FITS headers and verifying with proposal ID's on the HLA, but some may remain. 
\\
\\
From the above quality control, we reduced initial dataset from 280,105 individual exposures to 34,308 exposures that were used in the analysis. For each filter, we have 2468, 9332, 15844, 6664 exposures for the F435W, F606W, F814W and F850LP filters, respectively.
\\
\\
For each field, we constructed a database consisting of raw FITS header information and derived quantities. See \autoref{sec:fits_header_params} for a table of relevant FITS header keywords and StarView field names. This includes orbital information, time, date, exposure times, photometric calibration parameters, start and end date and times for every exposure, filter combinations for wheels 1 and 2, and field coordinates.

\subsection{Satellite Weather Data}\label{ceres_data}

Historical satellite weather data should be sensitive to the amount of Earthshine beneath a space telescope at a given time. In this work we utilise data from the ongoing CERES (Clouds and the Earth's Radiant Energy System) missions. CERES is a multi-satellite project dedicated to observing the Earth's global energy budget. The first satellite was launched in 1997, closely followed by Terra and Aqua in 1999 and 2002 respectively, in a Sun synchronous orbits. CERES still operates to the present time \citep{Doelling2013}, and CERES data is available for $60-70\%$ of HST's operational lifetime.
\begin{figure*}[ht!!!]
    \centering
    \includegraphics[width=11cm]{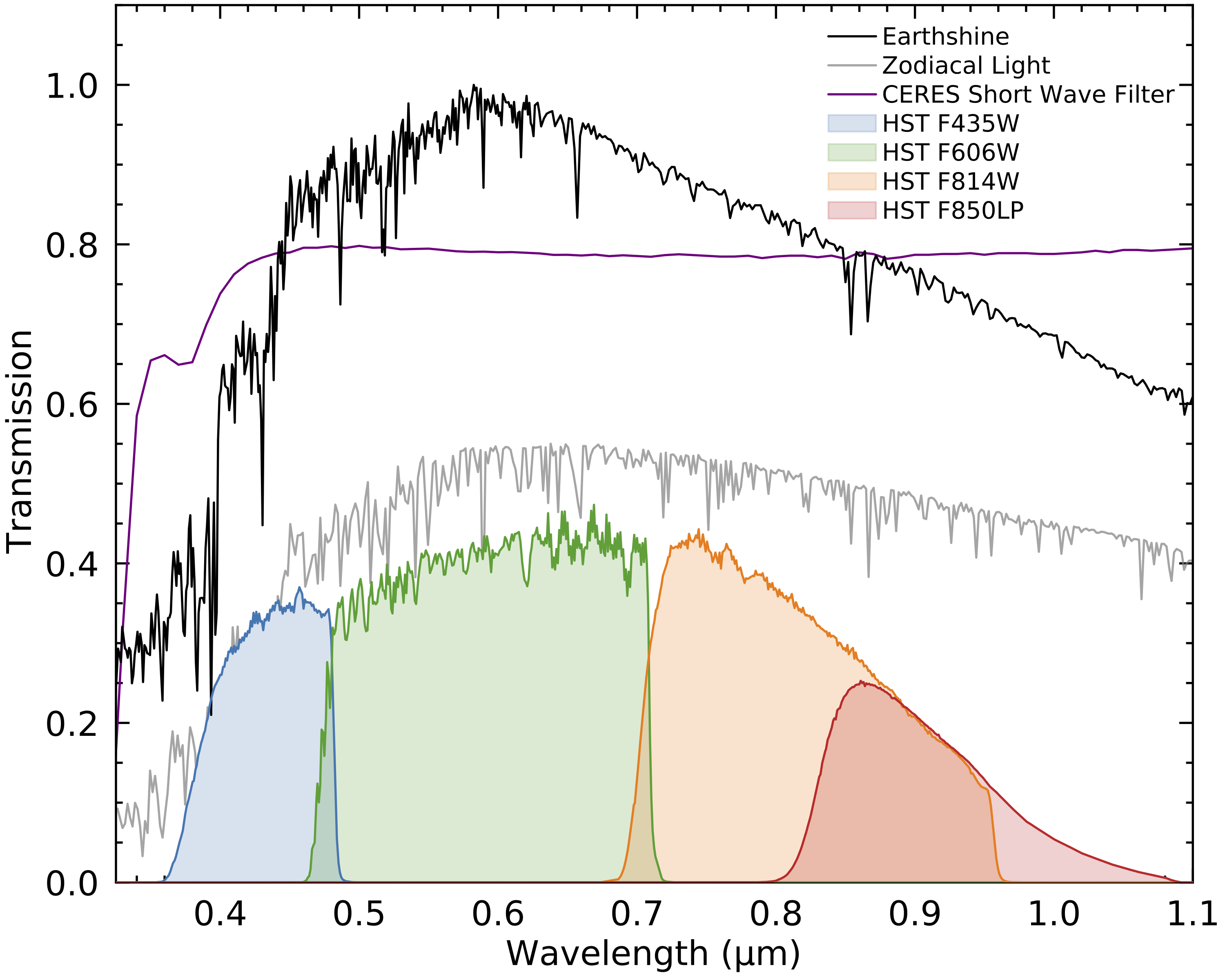}
    \caption{Filters used in this work plotted with respect to scaled spectra of components of the total sky. The Earthshine spectrum is approximated by a ASTM E-490 Solar Spectrum following \cite{Giavalisco2002} and is shown in black. The Zodiacal Light spectrum is shown in grey and described in \autoref{zl_mod}. Both spectra are shown with relative ``high'' scaling from \cite{Giavalisco2002} figure 2. Four HST filters are chosen for this study that span the visible wavelength portion of the scattered Earthshine and Zodiacal Light spectra. A portion of the CERES short wave filter from weather satellites, which extends to 5 $\mu$m, is shown in purple.}
    \label{fig:bandpasses}
\end{figure*}
\\
\\
CERES mission data is available online through NASA's Langley Research Center \footnote{\url{https://ceres.larc.nasa.gov/data/}}. We choose to focus only on the so called ``short wave''  CERES filter ($0.3 - 5 \mu$m) that most closely measures Earthshine levels in the chosen HST filters. An important caveat of interpreting analysis using the CERES data is that it is sensitive to thermal emission, while our HST filters will not be. CERES fluxes are total integrated flux measurements from the top of the atmosphere over a $1^{\circ}\times1^{\circ}$ latitude and longitude grid over a 3 hour period. The CERES shortwave filter, HST filters and the Earthshine and Zodiacal Light SED's are shown in \autoref{fig:bandpasses}. 
CERES shortwave instruments do not operate or are not sensitive to the night side of the Earth. As a result data products are only strictly relevant to HST acquired when it was over the daytime portion of the Earth. 
\begin{figure*}[ht!!!]
    \centering
    \includegraphics[width=11cm]{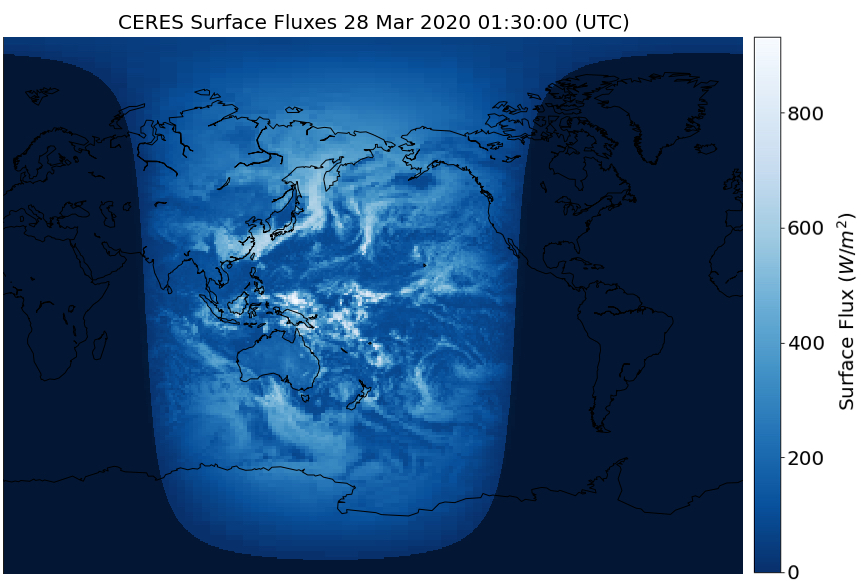}
    \caption{Earthshine observations by the CERES instruments are provided in $1^{\circ}\times1^{\circ}$ latitude and longitude grids with a temporal resolution of 3 hours on the illuminated side of the Earth. This figure illustrates one image from CERES projected onto a map of the Earth over Australia. Data used here are from the CERES short wave filter covering 0.3 - 5 $\mu$m.}
    \label{fig:figure4}
\end{figure*}
\\
\\
To compute the relevant CERES irradiance for a given observation, the primary scattering area (see \autoref{fig:angles_and_vectors}) of an exposure is projected onto the CERES spatial image of the entire illuminated Earth's surface. The average Earthshine flux inside the primary scattering area is calculated for the start, middle and end of each exposure. The calculated Earthshine in CERES native units of $W/m^2$ beneath the telescope for any one exposure is then the average of the 3 primary scattering area averages. For locations that have no CERES data (e.g. night-time side), the CERES image values are zero and will impact the averages in a way that accurately reflects the amount of Earthshine that potentially impacts an observation.
\\
\\
We calculate a primary scattering area beneath the telescope (see \autoref{fig:angles_and_vectors}) for the start, middle and end of every exposure. For this calculation, the location of the telescope needs to be known while an exposure is being acquired. The ephemeris for HST was retrieved from the North American Aerospace Defense Command Archive \footnote{\url{https://www.hsdl.org/c/tag/norad/}}. This includes data for HST over the past 20 years was kindly provided by private request. The ephemeris data was processed by the python package {\tt pyephem}\footnote{\url{https://rhodesmill.org/pyephem/}} to determine the location of HST at the start, middle and end of an exposure. The altitude of the telescope at the start of the exposure was adopted for the entire exposure.  The radius of primary scattering area is then given by: 
\begin{equation}
   R_{Earthshine} = \cos^{-1}\Big(\frac{R_{Earth}}{R_{Earth} + Alt}\Big) \times R_{Earth}
\end{equation}
For all calculations the radius of the Earth is assumed to be $6370 km$ and the exposures in this work have an average $R_{Earthshine}$ of $2535 km$.
\\
\\
A primary scattering area was computed for the location of HST at the start, middle and end of an exposure. The final primary scattering area for a given exposure was taken as the region covered by all 3 primary scattering areas. This tiling approach provides adequate overlapped sampling of the Earthshine beneath the telescope over the duration of the exposure.

\subsection{Training the Machine Learning Algorithm} \label{xgboost}

As will be shown in \autoref{isolate_stray_light}, basic trends between telescope attitude parameters, environmental parameters and the observed sky surface brightness can be visualised by sub-selecting the data set manually and utilising basic regression methods. However, due to the complexity and size of the data, we explored the use of a machine learning algorithm to capture the underlying trends in the data. {\tt XGBoost}\footnote{\url{https://xgboost.readthedocs.io/en/latest/}} is an efficient and flexible gradient boosting framework \citep{Chen2016} that can produce a multi-dimensional interpolation across the input parameters. 
\\
\\
We utilise the XGBoost algorithm to ingest and train on the observed sky surface brightness from HST images and corresponding observational parameters given in \autoref{sec:fits_header_params}. The model can then be used to predict the sky surface brightness for a desired location on the sky, time of year, and set of telescope attitude constraints (e.g. Sun Altitude, Limb Angle etc.) As described immediately below, the XGBoost algorithm can be trained in two different ways for each filter.
\\
\\
\textit{The single-field trained method}: This method focuses on identifying underlying trends in sky surface brightness towards a {\it single} HST field, made up of many pointings. It allows us to simplify the parameter space to investigate a single issue. For example, we use this method on GOODS North because the field has minimal Zodiacal Light, which allows us to better isolate the impact of stray light. Similarly we use this model to understand the benefit of the satellite weather data. The fields we use for this method are the Legacy HST fields, which provide a large amount of data spanning many years, and hence give us access to a fairly well-sampled orbital parameter space.
\\
\\
\textit{The all-sky trained method}: In contrast to the previous method, this method trains on all available data without any spatial restrictions. It is therefore sensitive to all components of the sky (e.g., stray light, Zodiacal Light and Galactic dust) and in principle should produce a model that can be used to predict the sky surface brightness for any field taken at any time of year. We will use this model to generate all sky maps of the sky as seen by HST.  
\\
\\
Despite the data quality control described above, there were a few spurious sky estimates that effectively contributed noise to the models. To mitigate the impact of outliers, for each of the 10 runs of the model we iteratively removed the upper 0.99 and lower 0.01 quantile in the residuals of the measured sky values and model predictions. 
\\
\\
Finally, to optimise performance, accuracy is favoured over speed for these tests. To prevent over fitting of the model early stopping rounds are set relatively high ($\sim 20 $). The learning rate is tuned to a low value of 0.01, and we allow the model a large number of estimators at 10000. Computation time can be sped up considerably with fewer estimators, however this comes at the cost of accuracy. 
\\
\\
The software developed for this work including worked examples, software versions, calibration data and example data are available on the {\tt Skypy} \footnote{\url{https://github.com/Physarah/skypy}} github repository.

%% file: 03_results.tex
\section{Results}
\begin{figure*}[ht!!!]
    \centering
    \includegraphics[width=11cm]{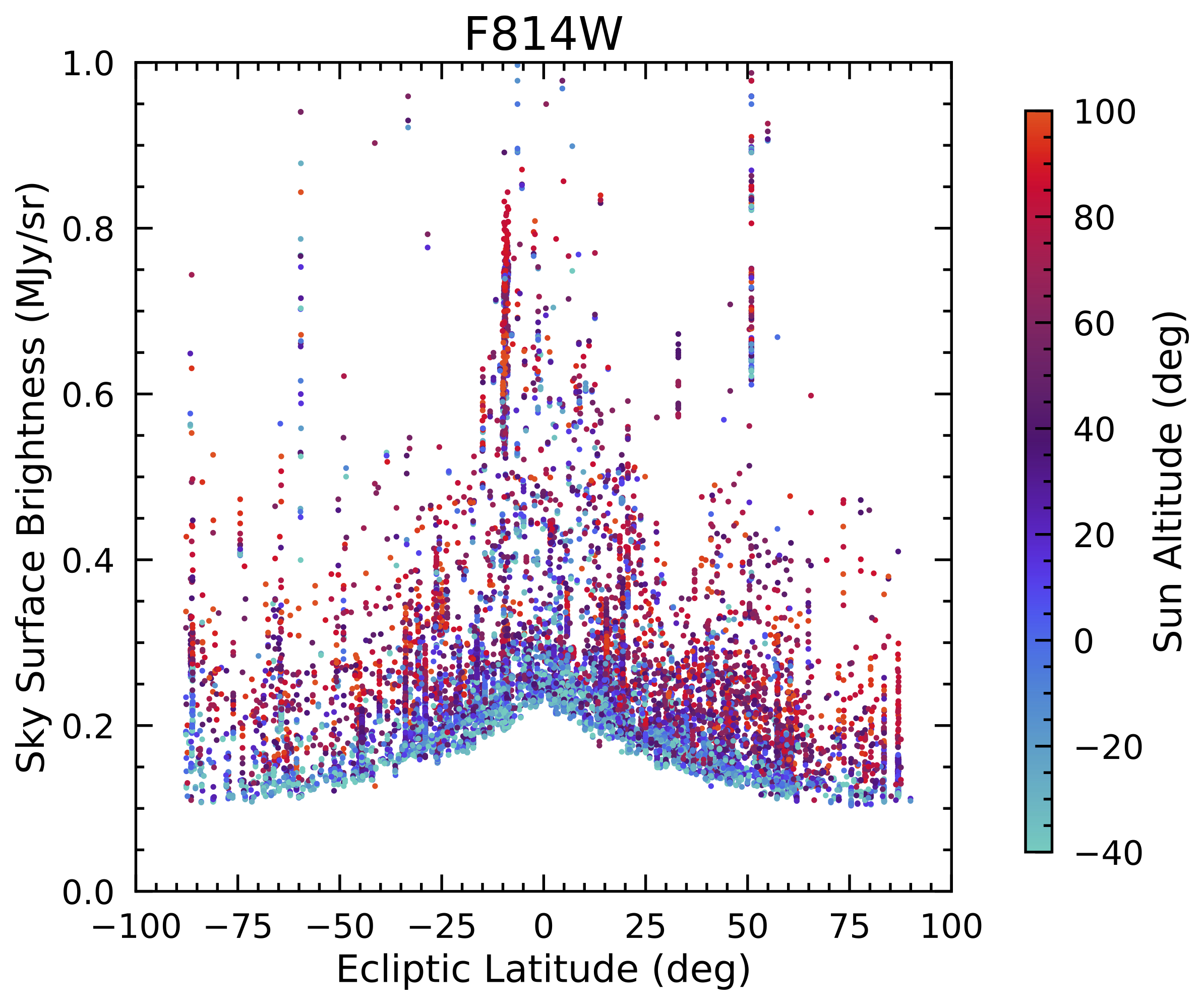}
    \caption{Total observed sky is plotted as a function of ecliptic latitude to illustrate the change in minimum sky backgrounds due to the planar distribution of Zodiacal dust within the Solar System. Scatter upwards from the minimum backgrounds shows some correlation with Sun Altitude, indicating a likely increased contribution from Earthshine stray light. However, the correlation with Sun Altitude is not perfect, indicating other underlying parameters impact the observed sky. Single fields observed over a range of Sun Altitudes and/or Zodiacal levels are apparent as vertical stripes.}
    \label{fig:figure1}
\end{figure*}

A subset of the data used in the analysis is plotted as a function of ecliptic latitude for the F814W filter in \autoref{fig:figure1} to motivate how we approach the analysis of the data. The figure illustrates some key physical phenomena relevant to this data set. Data points are coloured according to the mean Sun Altitude at which the exposure is taken. The sky scatter follows a “Christmas Tree” shape defined by ecliptic latitude
\\
\\
We expect multiple parameters to determine the time-variable sky surface brightness in any field observed from LEO. For example, \autoref{fig:figure1}, shows an increase in minimum sky surface brightness for fields close to the ecliptic plane. This is most likely due to the contribution of Zodiacal Light to the total sky. 
\\
\\
The lowest sky for any ecliptic latitude occurs when the Sun Altitude is less than $0^{\circ}$, which corresponds to the night side of an orbit. This happens because exposures should be minimally impacted by Earthshine stray light in the Earth's shadow at $~\sim 0.8\mu$m.
\\
\\
We also note there is evidence in the plot for other parameters contributing to the sky surface brightness. For instance, at Sun Altitude $> 0^{\circ}$, the sky surface brightness is typically elevated and can reach values $\sim 4\times$ the lowest observable sky.  Furthermore we note that fields with significantly elevated sky surface brightness but low Sun Altitude are likely due to Galactic cirrus contributions or poorly calculated sky measurements in crowded fields.
\\
\\
To conclude, \autoref{fig:figure1} shows that Zodiacal Light and Sun Altitude are both important factors in determining the sky. The reason Sun Altitude is important is also reflected in \autoref{fig:figure4}. At the center of the illuminated portion of the Earth, there is a higher amount of reflection from the Sun due to the more optimal reflection geometry. This is why Sun Altitude is a good first-order indicator of the potential for Earthshine stray light to impact the sky. Thus given that Earthshine and Zodiacal Light appear to dominate the sky, we therefore characterise their impact in the next section before exploring other parameters in \autoref{fig:relationships}.

\subsection{Exploring the dominant components of the sky}\label{zl_mod}

\begin{figure*}[ht!!!]
    \centering
    \includegraphics[width=18cm]{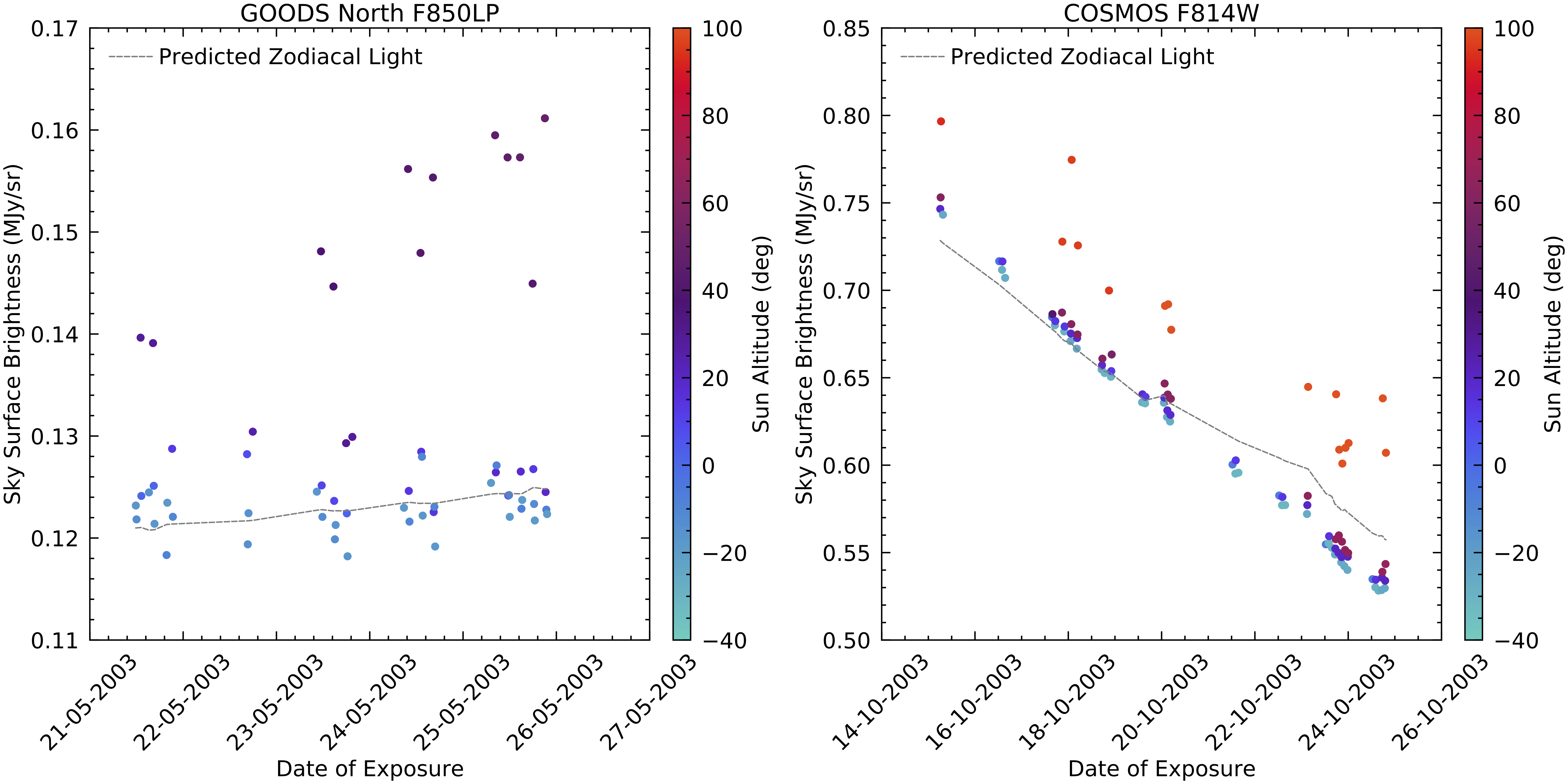}
    \caption{The observed sky surface brightness in the GOODS North fields, and the COSMOS fields in F850LP and F814W filters, respectively. Observations are coloured by the mean Sun Altitude at which the observation was taken. The predicted Zodiacal Light is shown in grey. For the GOODS North field on the left panel, which is high off the ecliptic plane, data points scatter upward from the baseline Zodiacal models by as much as $34\%$ of the total observed sky surface brightness. These data points correspond to observations with high Sun Altitudes, which is consistent with the idea that increased Earthshine is able to get into the telescope at higher Sun Altitudes (see \autoref{fig:angles_and_vectors}). For the COSMOS field on the right panel, we see the data approximately follows the predicted Zodiacal Light variation with time. Scatter to higher sky surface brightness reflects various observed Sun Altitudes on approximately the same date.}
    \label{fig:time_variability}
\end{figure*}

The impact of the two most dominant components of the sky - Earthshine and Zodiacal Light - in HST fields can be isolated by considering fields in two different locations. Close to the ecliptic plane the sky will tend to reflect how the Zodiacal Light varies in time due to the motion of the Earth around the Sun caused by changes in viewing angles through the non-homogeneous Zodiacal cloud over the period of a year \citep{Leinert1998}. A field far from the ecliptic plane will have minimal Zodiacal Light that also does not vary significantly. This enables us to separate the impact of Earthshine due to changing viewing angles and variable weather, from changes due to Zodiacal Light. 
\\
\\
To help isolate time-varying stray light contributions from time-varying Zodiacal Light intensities, we use Zodiacal models to help us identify when a change is dominated by a changing Zodiacal Light intensity. For this study we use the Zodiacal model described in \cite{Leinert1998}, with amendments by \cite{Aldering2002} and calibrated using fluxes from the GOODS North field \citep{Giavalisco2002}. This method is favoured over the Wright and Kelsall models as it can be easily adapted to the shorter wavelengths used in this study. It should be noted that the Zodiacal Light model is not used in the process of the empirical analysis described in \autoref{isolate_stray_light} or XGBoost models described in \autoref{model_performance}. This ensures that the results unbiased by the assumptions that are typically used to construct Zodiacal models. 
\\
\\
To implement the \cite{Aldering2002} Zodiacal model, we adopted the relative intensity of the Zodiacal Light with respect to the north ecliptic poles as calculated in {\tt Gunagala}\footnote{\url{https://github.com/AstroHuntsman/gunagala}}. This follows the prescription of \cite{Leinert1998} by interpolating the results of their table 17. The absolute intensity of the Zodiacal Light for each of the HST filters is scaled using the observed sky surface brightness found at the the north ecliptic poles given in \cite{Giavalisco2002}. Following \cite{Aldering2002}, the resulting intensity is multiplied by a correction factor of 0.49 for $\lambda > 0.5 \mu$m and 0.9 for $\lambda < 0.5 \mu$m to account for the reddening of the Solar spectrum in scattered Zodiacal Light.
\\
\\
As shown in \autoref{fig:time_variability}, we use COSMOS F814W observations to probe the ecliptic plane and compare it to GOODS North F850LP observations which is high off the ecliptic plane. While this comparison uses two different filters, we expect this has a negligible impact on the results because the filters (see \autoref{fig:bandpasses}) cover a similar portion of the Zodiacal Light and Earthshine Spectral Energy Distributions (SEDs).
\\
\\
\noindent
In \autoref{fig:time_variability}, the expected Zodiacal Light contribution to the sky is shown for the same time and location on sky that each exposure is taken. The sky surface brightness is shown coloured by the Sun Altitude recorded at the midpoint of the exposure. In the GOODS North field, most exposures have a sky surface brightness close to the Zodiacal Light model estimate $ \sim 0.12$ MJy/sr over the 5 day period. However, some reach as high as $ \sim 0.16$ MJy/sr. This difference is almost $\sim 34 \%$ of the expected Zodiacal Light intensity.
\\
\\
In the COSMOS field at the start of the observation period, the lowest sky surface brightness is $ \sim 0.75$ MJy/sr with deviations reaching $\sim0.80$ MJy/sr, or $\sim 7\%$ of the mean sky surface brightness at the beginning of this period. At the end of the 10 day period, the lowest sky surface brightness drops to $\sim0.52$ MJy/sr with a maximum of $\sim0.65$ MJy/sr, or $\sim23\%$ of the lowest sky surface brightness at the end of the observing period.
\\
\\
These results suggest that there are two main time-variable components of the observed sky from LEO in the $\sim 0.8 - 0.9 \mu$m range. One component corresponds to a gradual change as a function of time on the order of days, and is seen to follow the predicted Zodiacal Light estimate. The second component changes on the time scale of hours, and is seen to correlate with the Sun Altitude of the exposure. For fields close to the ecliptic pole, the effect of Sun Altitude variations is most apparent - with the time-variable sky changing by $\sim 34\%$ of the predicted Zodiacal Light on the time scale of hours. 
\\
\\
In this section, we have shown that the data have two major time-varying components, which we show are likely due to varying Zodiacal Light levels and a component correlated with the Sun Altitudes of a particular exposure. As discussed earlier, Sun Altitude is likely a measure of the potential of stray light from Earthshine, thus we have demonstrated that even within a single orbit, HST backgrounds can vary by as much as $\sim30\%$ due to stray light from Earthshine.

\subsection{Isolating the stray light component}\label{isolate_stray_light}

In \autoref{fig:time_variability} we isolated a time-variable component of the sky that is not due to the change in Zodiacal Light intensity, and is correlated with the mean Sun Altitude at which the exposure was taken. This result suggests that orientation of the telescope with respect to sources of stray light is an important factor in understanding the surface brightness of the sky.
\\
\\
Here we attempt to isolate key parameters available to us in HST FITS headers that correlate most strongly with sky surface brightness and may aid in characterising the stray light contribution. This is done by projecting the data into two dimensions to illustrate how each parameter defined in \autoref{fig:angles_and_vectors} impacts the sky surface brightness. Here we use only use exposures from the GOODS North fields. These fields have high ecliptic latitudes and effectively avoid complications from the time-varying Zodiacal Light component. 
\begin{figure*}[ht!!!]
    \centering
    \includegraphics[width=17cm]{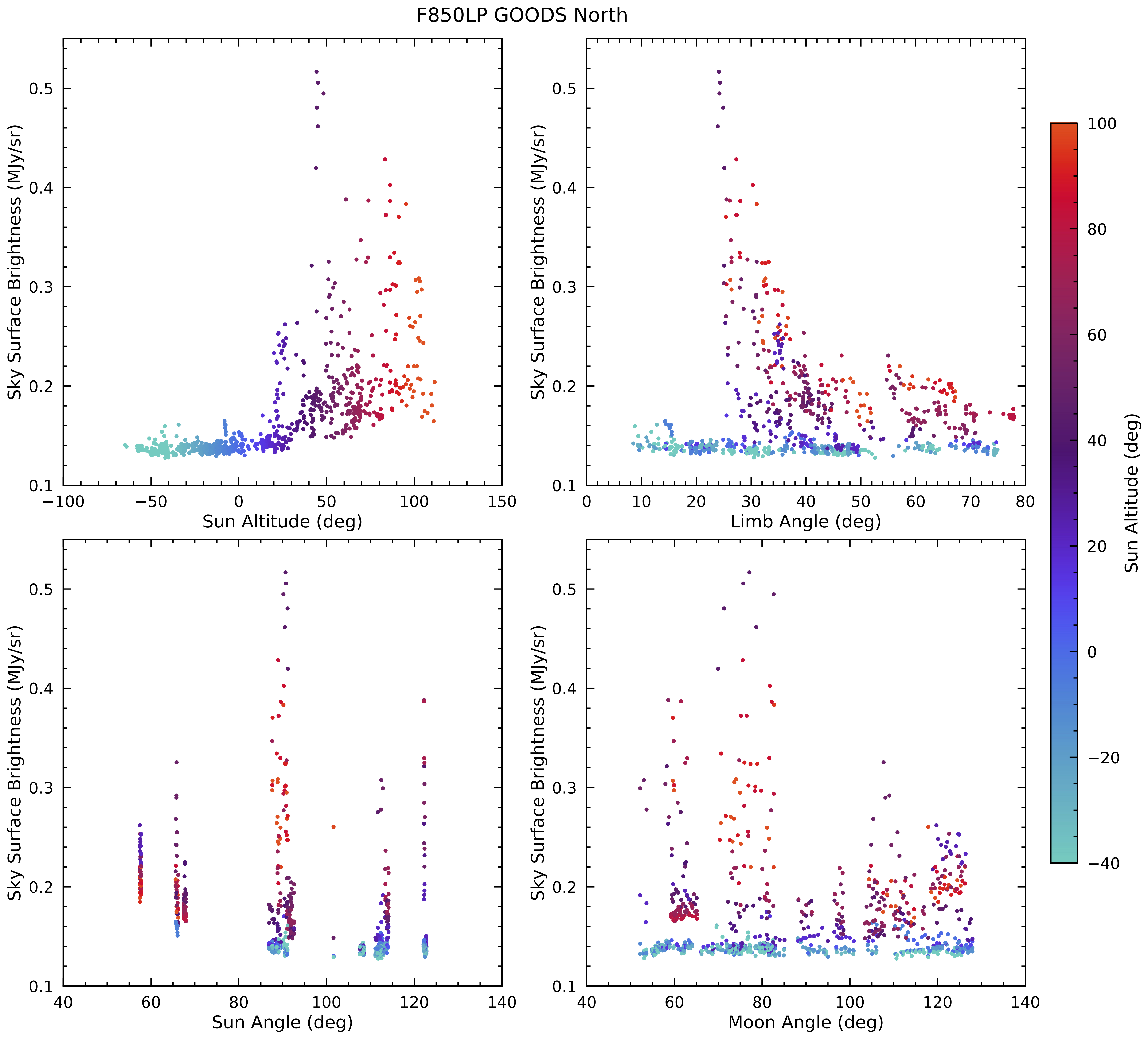}
    \caption{Observed HST F850LP sky surface brightness from observations in the GOODS North fields against various telescope attitude angles. In order to assess the sensitivity of the parameters to stray light, we selected a field with high ecliptic latitudes and minimal time-variable Zodiacal Light contribution. {\it Top left panel:} Sun Altitude shows a strong correlation with the sky surface brightness. This indicates observations taken at nighttime (i.e. Sun Altitude $< 0^{\circ}$) lead to lower measured sky, presumably due to minimal stray light from Earthshine. Observations over the daytime (i.e. Sun Altitude $> 0^{\circ}$) portion of the Earth shows a higher sky surface brightness likely from increased Earthshine stay light contributions. {\it Top right panel:} shows sky surface brightness as a function of Limb Angle. Observations over a daytime Earth show a correlation with Limb Angle, indicating the angular distance between a target field and the Earth impacts the amount of Earthshine stray light entering the telescope. For nighttime observations taken in the Earth's shadow there appears to be little correlation until very low Limb Angles of $<20^{\circ}$) are reached.
    {\it Lower panels:} Sun Angles below $\sim80$ and Moon Angles below $\sim70^{\circ}$ also show correlations as sky surface brightness increases as HST points closer to the Sun and Moon, respectively. The variation in the minimum observed sky is likely due to the variations in Zodiacal Light intensity with time.}
    \label{fig:relationships}
\end{figure*}
\\
\\
\autoref{fig:relationships} shows the sky surface brightness data against various orientation angles from F850LP observations at a high ecliptic latitude. In the top left panel of this figure, it is evident that Sun Altitude has a correlation with the sky surface brightness and illustrates the idea that stray light from Earthshine is significantly more important when HST is observing over a portion of the daytime Earth (i.e. Sun Altitude $>0^{\circ}$). At low Sun Altitudes HST will not only be over a nighttime portion of the Earth, it will also likely be in the Earth's shadow, further reducing possible stray light from Sunshine.
\\
\\
The top right panel of \autoref{fig:relationships} shows a bifurcation at low Limb Angles, which reflects the underlying trend with Sun Altitude. For the data taken when HST was over the daytime Earth, there is a correlation with Limb Angle. This likely illustrates that more stray light from Earthshine can enter the telescope when the field is close to the Earth's limb from the perspective of HST. This is consistent with preliminary work by \cite{Biretta2003} who report that Limb Angles $<25^{\circ}$ can show significantly elevated sky levels. We also note that for exposures of Sun Altitude $>0^{\circ}$, to protect the instrumentation from potential damage, Limb Angles are restricted to $>20^{\circ}$.
\\
\\
The lower left panel of \autoref{fig:relationships} shows that Sun Angle, which is determined by the position of the Earth with respect to its orbit around the Sun, is discretised into groups. These groups roughly map onto HST ``visits'' of the GOODS North field over time, at varying Sun Angles. For Sun Angles $>80^{\circ}$, the minimum sky surface brightness increases. This could be explained by an increase in Sunlight entering directly into HST and/or an increased Zodiacal Light contribution because HST is looking closer to the Sun. The vertical stripes of data correspond to a single visit and illustrate an increase in sky surface brightness due to the various combinations of viewing angles. 
\\
\\
The impact of scattered light from the surface of the Moon is explored in the lower right panel of \autoref{fig:relationships}. To minimise the impact of lunar scattered light, HST is restricted to Moon Angles greater than $20^{\circ}$. No evident trend is seen in these data, which is limited to Moon Angles $> 50^{\circ}$. This could reflect the fact that Moon Angle restrictions of $> 50^{\circ}$ works well. Or it might reflect the relatively sparse temporal sampling of the current dataset, which would limit the range of lunations this analysis can probe.
\\
\\
The data presented in \autoref{fig:relationships} indicates that it is possible to minimise the contribution of stray light in HST data by limiting the range of angles in which the telescope can operate {\citep[e.g.][]{Biretta2003}}. We note that some data do not follow the trends outlined in the following analysis likely because of limitations in the sample quality control described in \autoref{data_quality_control}. The top left panel of \autoref{fig:relationships} shows minimal scatter for Sun Altitudes $\sim < -10^{\circ}$. Employing this Sun Altitude selection to the data in the top right panel of \autoref{fig:relationships} we find that scatter is minimal for Limb Angles $ \sim > 20^{\circ}$. With the same Sun Altitude selection, we find that Sun Angle should be limited to be $ > 80^{\circ}$ to minimise the scatter. And finally, we find no strong evidence for a need to select on Moon Angle to minimise scatter in the sky observations. 
\\
\\
In summary, the data presented in \autoref{fig:relationships} gives some indication of what telescope attitude parameters contributes most significantly to the F850LP sky in the direction of the North ecliptic pole. Maximum sky levels are observed when HST is over the daytime sky and when HST is pointing at a field with a low angular distance to the Earth. We also find angular distance to the Sun is a secondary factor, at least in the limited dataset that was used. Note also we cannot separate the impact of increased Zodiacal Light intensity with the possibility of direct Sunlight scattering into the telescope aperture. The Moon does not appear to significantly impact the sky surface brightness, though this may only reflect the existing HST Moon Angle viewing restrictions and/or a limited range of lunations in the current dataset.
\\
\\
\subsection{Using Satellite Weather Data to Improve Earthshine contribution estimates}\label{satellite}

In \autoref{the_start} we raised the idea of improving stray light predictions from Earthshine levels using satellite weather data of the Earth. Here we investigate the possibility of utilising satellite weather data from CERES to see if it can help identify exposures with increased Earthshine levels in HST data. 
\begin{figure*}[ht!!!]
    \centering
    \includegraphics[width=11cm]{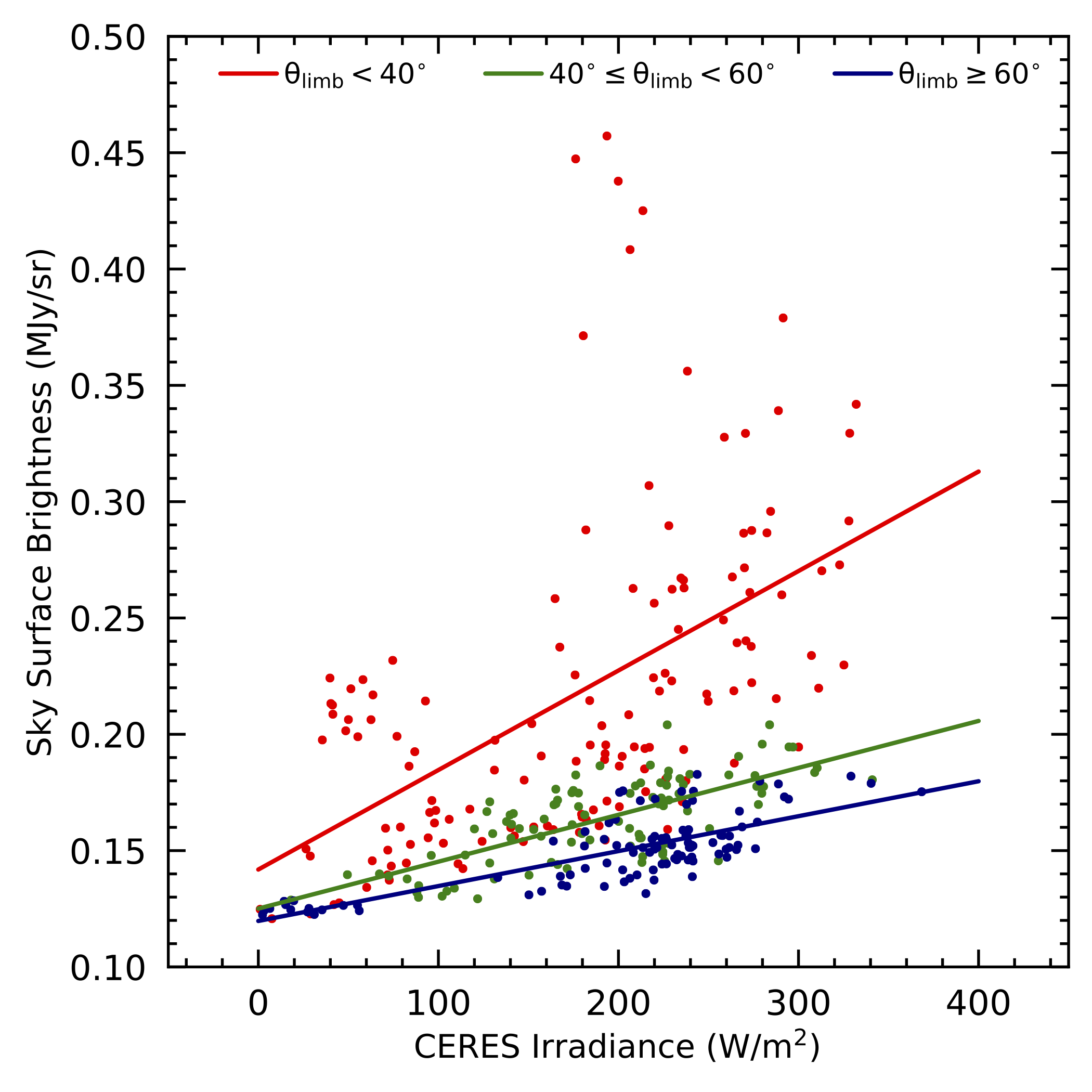}
    \caption{Sky surface brightness versus total irradiance from the space-based CERES weather data. Sky surface brightness data is from the F850LP filter in GOODS North field.  CERES irradiance is the sum of CERES flux from observations in the estimated region on the Earth directly below HST during an exposure. The data generally show a correlation, indicating that the specific weather or region on the Earth below Hubble at least partially determines the sky surface brightness levels. We note this is direct evidence that stray light from Earthshine can significantly impact HST sky surface brightness estimates. The data show significant scatter, especially at higher CERES fluxes. Colours show 3 different Limb Angle bins with lines fit to each to highlight an underlying correlation. Higher Limb Angles result in a shallower relationship, likely because HST is pointing further away from the Earth and hence will receive less stray light from Earthshine.}
    \label{fig:ceres_flux}
\end{figure*}
\\
\\
\noindent
\autoref{fig:ceres_flux} shows the HST sky surface brightness levels against the CERES irradiance. As detailed in \autoref{ceres_data}, the irradiance is the average of flux within the primary scattering area, which is the estimated area below HST during an observations (see \autoref{fig:angles_and_vectors}). As in \autoref{isolate_stray_light}, in this analysis we use data for the F850LP filter in the GOODS North field to minimise the impact of a time-variable Zodiacal Light component. As discussed in \autoref{ceres_data} the CERES data is only given on the daytime side of the Earth. For this reason, this analysis is limited to exposures with a Sun Altitude $> 0^{\circ}$. 
\\
\\
The data in \autoref{fig:ceres_flux} show a trend where higher CERES levels generally correspond to higher HST sky surface brightnesses. This confirms that Earthshine indeed can dominate HST sky surface brightnesses in certain orbital configurations. To explore the cause of the scatter in the relationship between CERES irradiance and HST Sky surface brightness, the data in \autoref{fig:ceres_flux} are divided into 3 Limb Angle bins as indicated in the figure. Not surprisingly, higher sky surface brightness levels are found at fixed CERES irradiance due to the fact the telescope is pointing closer to the Earth at smaller Limb Angles. 
\\
\\
To conclude we show that data from the Earth observing satellites CERES does appear to be useful in identifying exposures with higher probability of elevated background due to stray light from Earthshine. 

\subsection{The XGBoost model of the sky}\label{model_performance}

\begin{figure*}[ht!!!]
    \centering
    \includegraphics[width=18cm]{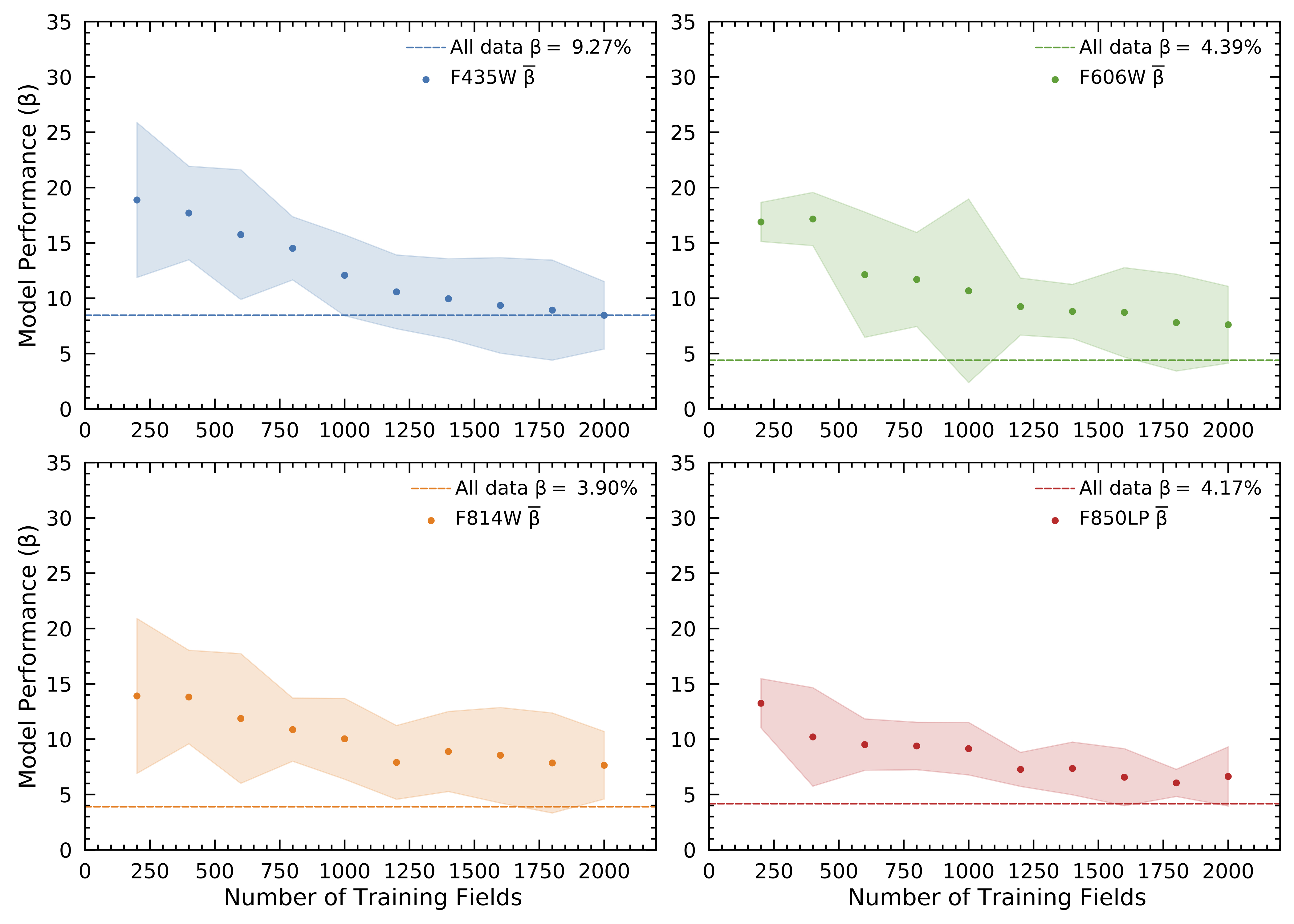}
    \caption{Assessing the performance of the XGBoost machine learning model (see \autoref{xgboost}) for various training sample sizes. Model performance is the mean percentage difference in flux between the measured and predicted sky. The mean percentage difference is calculated from 10 training runs each using a random training sample of the size indicated. The shaded regions reflect the standard deviation of the 10 training runs. Horizontal line in each panel is the model performance when the maximum sample size is used. By a sample size of 2000, the filters from each model have approximately the same model performance of $\sim10\%$. When the maximum sample size is used, the best-performing filter is F814W with a performance average of $3.90\%$, reflecting the filters largest sample size. This is also an indication that performance can be improved even further with larger training set.}
    \label{fig:test_train}
\end{figure*}

\begin{figure*}[ht!!!]
    \centering
    \includegraphics[width=18cm]{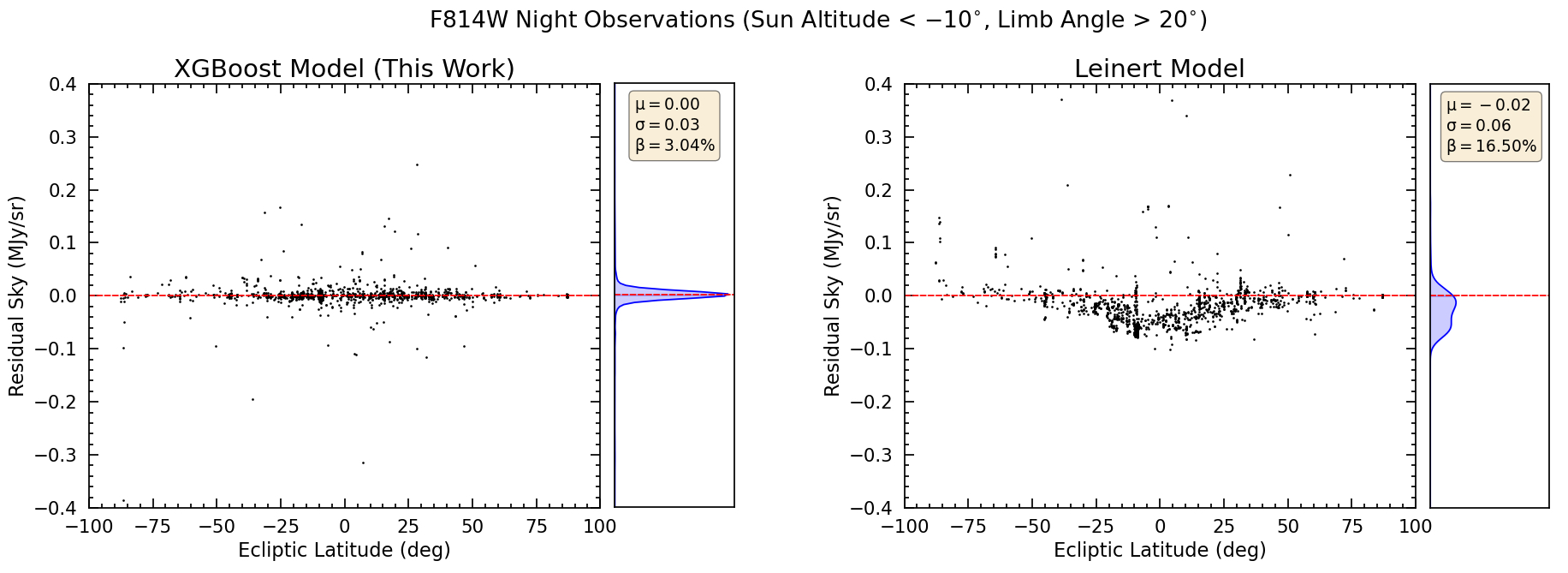}
    \caption{Residuals for the predicted sky from the Leinert, and XGBoost models subtracted from observed F814W night time sky. Data were restricted to minimise stray light contributions using Sun Altitude $< -10^{\circ}$ and Limb Angle $> 20^{\circ}$. Blue distributions to the right of each figure shows the relative smoothed kernel density estimation distribution of the residuals. The XGBoost model is able to predict the sky to within an average of $\beta$ of $3.08 \%$ of the true sky. The Leinert shows offsets where the model has over estimated the observed sky near the ecliptic plane. This offset may reflect inherent biases in the model's interpretation of the structure of the inner Solar system interplanetary dust cloud.}
    \label{fig:model_comparisons}
\end{figure*}

In the above analysis we have isolated the Zodiacal Light and stray light time-variable components of the sky by restricting analysis to single fields, and breaking the problem down into a series of two dimensional relationships with key telescope attitude parameters. In doing so we have identifying initial constrains on telescope attitude to limit the impact of stray light shown in \autoref{isolate_stray_light}. We also identify a new parameter - the direct observed Earthshine beneath a space telescope - that is useful in identifying exposures contaminated by stray light. However if this analysis is to be extended to larger datasets in multiple fields, multiple wavelengths, and more extensive telescope attitude parameter spaces, a more efficient method of exploring the data is needed.
\\
\\
As detailed in \autoref{xgboost}, we constructed a machine learning model based on the XGBoost algorithm to predict sky surface brightness levels for any set of telescope attitude parameters and locations on the sky. In the following series of tests, we explore how well the model can predict the brightness of the sky in any field observed with a particular set of telescope attitude parameters. To assess the models performance, we define a parameter ($\beta$) in units of percentages that describes how well the model can predict the sky surface brightness of a test field. Model performance is the mean percentage difference in flux between the measured and predicted sky as defined below:
\begin{equation}
     \beta = \frac{\sum \big( \frac{|S_{test} - S_{pred}|}{S_{test}} \big) \times 100\%}{N}
\end{equation}
Where N is the size of the test dataset, $S_{test}$ is the test sky surface brightness measurement, and $S_{pred}$ is the predicted sky estimate from the model.
\\
\\
Firstly we explored the impact of sample size on the performance to ensure filters with different numbers of test fields are compared equally. This also allowed us to explore the possibility for wavelength dependent variation that may exist in the data, by training the model with data from 4 different filters. The model is trained using the all-sky trained method on a randomly selected subset of data from 200 - 2000 exposures across the entire sky. The test data used to asses the performance of the model is randomly selected each time from a 70:30 split of testing and training data. The performance per training field is calculated by taking the average of 10 individual training runs, with randomly selected training samples to ensure e.g. single spurious pointings do not dominate the results and also as a way to estimate uncertainties on the model output. In the following the average performance and associated standard deviation of the 10 repeated training outputs is reported.
\\
\\
\autoref{fig:test_train} shows how the model performance improves with larger training sample sizes. For these initial tests, no CERES data is used nor any Zodiacal Model input. We note the performance of each filter is approximately the same when the sample size is similar, suggesting that there is no apparent wavelength dependence of the models performance.
\\
\\
We then utilise all available data for each filter to train and test. The performance of each model is found to be better than $10\%$ for all filters, with F814W performing the best at an average of $3.5 \pm 0.1\%$, followed by F850LP at $3.7 \pm 0.2\%$, F606W at $4.0 \pm 0.2\%$, and F435W at $8.0 \pm 0.9\%$. The improvement in performance appears to reflect the training sample size for each filter. These results demonstrate the ability of the model to predict the sky surface brightness of a particular field given a set of unique telescope attitude parameters to well within the $\sim 34\%$ variation in sky surface brightness found in \autoref{zl_mod} due to the presence of stray light. 
\\
\\
In \autoref{satellite} we found that satellite weather data contains some information that might be used to better predict the sky surface brightness of a particular field. To explore how folding in CERES data to the model changes the performance, we train the model on the GOODS North field with the F850LP filter with and without calculated CERES irradience. We find a change in performance of only $0.1 \pm 0.3 \%$ which is within the limit of error of the model. Uncertainties on this change in performance were estimated by running the training 10 times using 10 randomly selected training datasets. While this may indicate that CERES data does not significantly improve the performance of the machine learning model, it is somewhat surprising given the evident correlation between the observed sky and the CERES data in \autoref{satellite}. A perhaps relevant complication with the CERES data is that it covers $0.3-5\mu$m, so it will contain thermal emission missed by the F850LP filter. More CERES fluxes calculated for a greater range of HST fields, wavelengths and observing conditions may help to understand the contribution of CERES data to the machine learning model. 
To evaluate how the model performs in comparison to other existing Zodiacal Light models, we calculate the expected Zodiacal Light intensity from the {\tt{Gunagala}} implementation of the Leinert model described in \autoref{zl_mod}. The Leinert model does not contain an estimate for the contribution of stray light, so to ensure a fair comparison, we restrict our sample to night time data only with Sun Altitudes $< -10^{\circ}$ and Limb Angles $ > 20^{\circ}$. We use F814W as it is the largest subset of the data presented in this work, and choose to use fields across the entire sky to sample a large range of ecliptic latitudes. 
\\
\\
The residual sky is calculated by subtracting the model sky from the observed sky and plotted as a function of ecliptic latitude in \autoref{fig:model_comparisons}. The XGBoost model presented in this work achieves a Gaussian-like distribution about a mean residual of 0, a standard deviation of 0.03 MJy/sr and a $\beta$ of $3.04 \%$. We note this is slightly lower than the value determined in \autoref{fig:test_train} as the training and testing data set only contains night time data. In comparison, the Leinert model has a mean residual of $- 0.02$ MJy/sr, a standard deviation of 0.06 MJy/sr and a $\beta$ of $16.50 \%$. The Leinert model residuals contain an offset that varies with ecliptic latitude. This offset is largest when fields are located close to the ecliptic plane, and may indicate that the model is not capturing the complex structure of the interplanetary dust cloud in the inner Solar system \cite{Korngut2021, Jorgensen2021}.
\\
\\
In summary, the results of \autoref{model_performance} highlights the benefit of an empirically generated sky model. The model is able to perform as well as, if not better, than existing analytical/physical models of the observed sky. It is flexible and can be trained on single or multiple fields, can be easily modified to incorporate new training parameters to improve model accuracy, and performs equally well across F435W, F606W, F814W and F850LP for similar sized training datasets. The model is also able to maintain its high level of accuracy when stray light impacted fields are introduced, demonstrating that it may be valuable tool to predict the observed sky brightness under a range of different observing conditions.

%% file: 04_discussion.tex
\section{Discussion}\label{sec:discusssion}
\begin{figure*}[ht!!!]
    \centering
    \includegraphics[width=11cm]{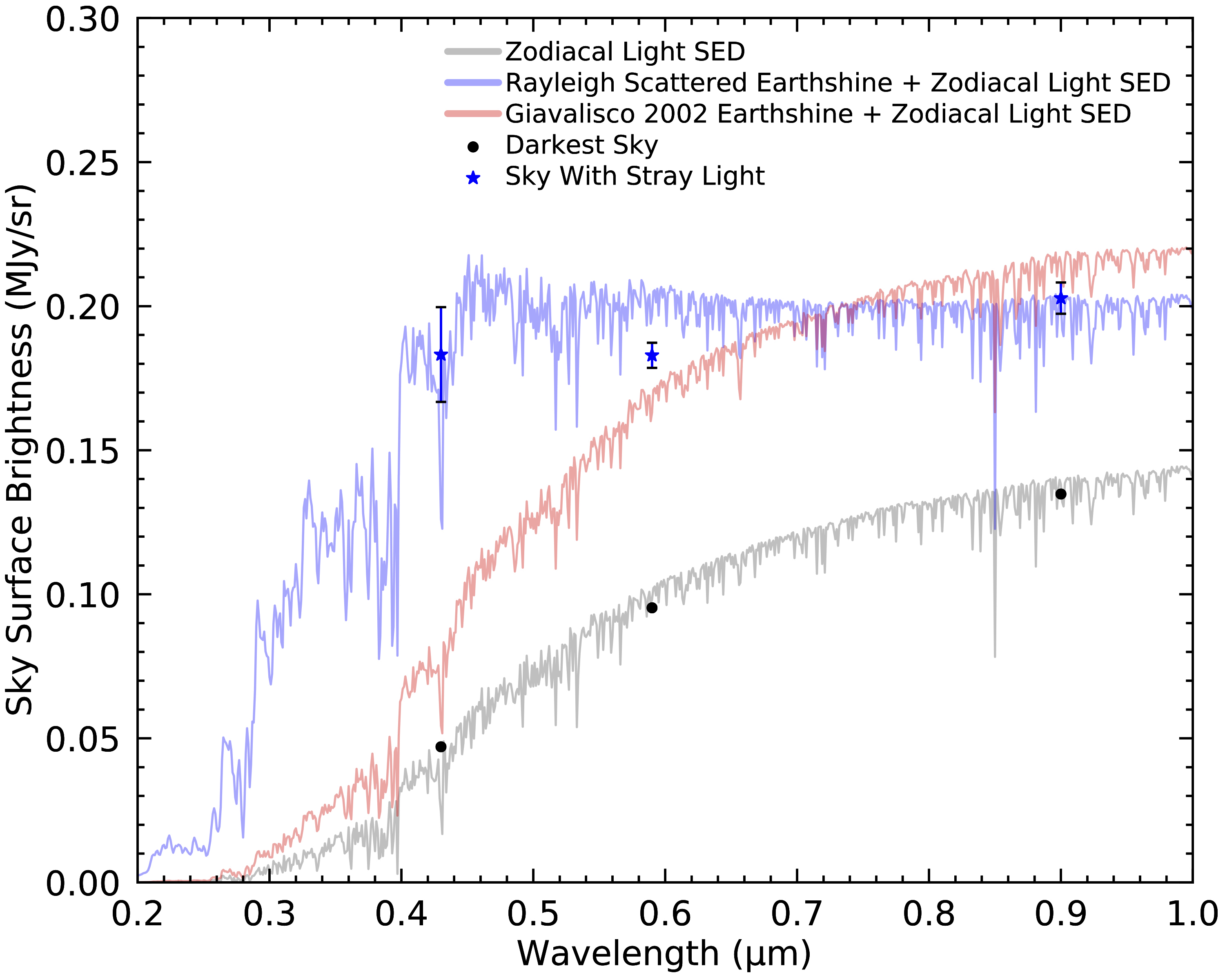}
    \caption{The SEDs of averaged sky measurements from fields that were selected using criteria given in \autoref{isolate_stray_light} to have minimal stray light as well as fields selected to have a larger stray light contribution. The standard deviation, normalise by the sample size of the observations are also plotted as error bars on all data points. The normalised standard deviation of sky surface brightness for fields with higher stray light contribution is larger as these exposures span a larger range of Sun Altitudes and Limb Angles to ensure a larger sample size. The SED from the fields selected to have minimal stray light follow the \cite{Aldering2002} predicted Zodiacal Light SED model towards the GOODS North fields. The SED of the fields with stray light cannot be fitted with a Zodiacal Light, Earthshine or Solar spectrum alone. A Solar plus Zodiacal spectrum also did not fit the relatively blue observed SED. To mimic the spectrum of stray light from Earthshine, we combined a Rayleigh scattered Solar spectrum component with a Zodiacal Light spectrum and fit these to the data. This model is able to reproduce the F435W excess seen in the stray light contaminated data. }
    \label{fig:sed_test}
\end{figure*}

\begin{figure*}[ht!!!]
    \centering
    \includegraphics[width=22cm, angle=90]{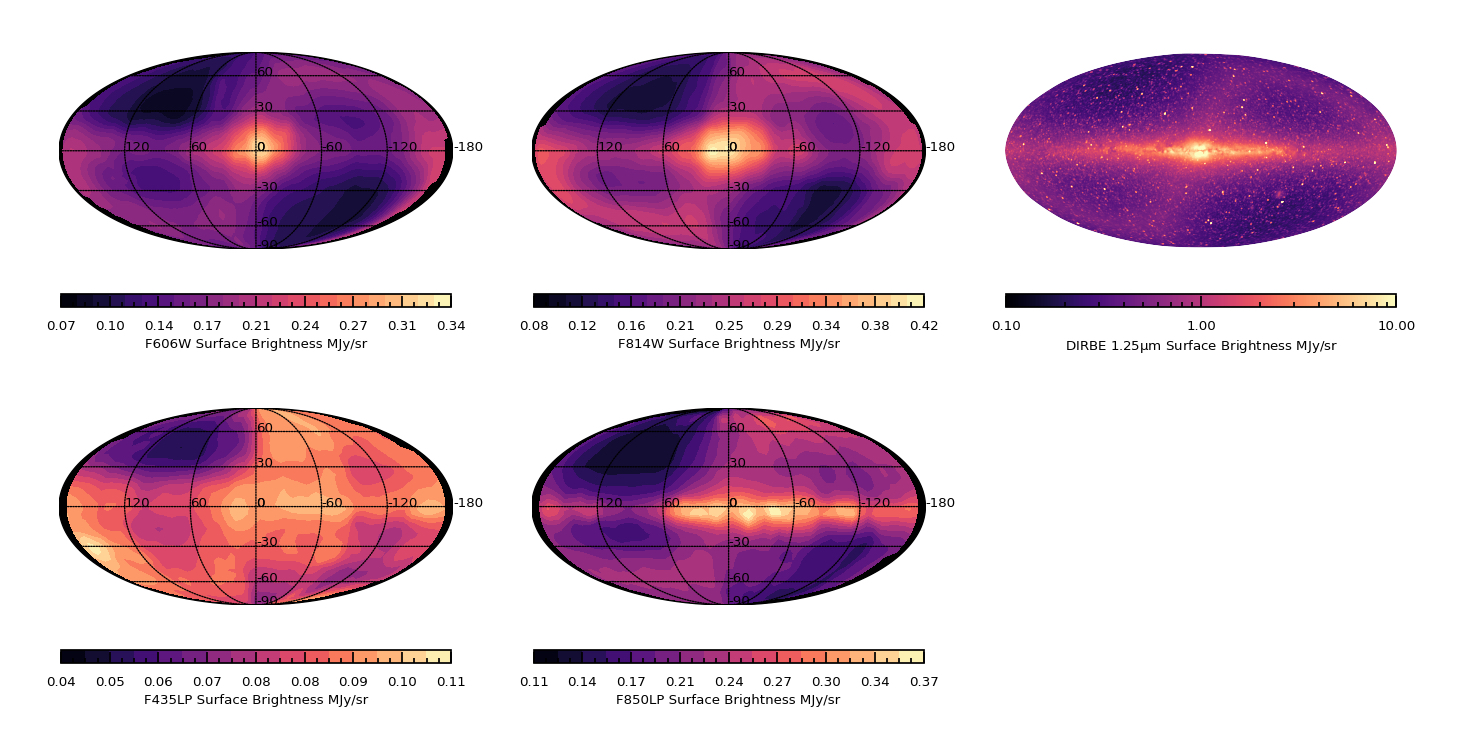}
    \caption{All sky maps in Galactic coordinates from sampling our sky model. The models are trained on the entire data set, and are used to predicted the sky for a set of constraints that we expect to have minimal contributions from Earthshine, Moonshine and Sunshine (constraints given in \autoref{isolate_stray_light}). The sky surface brightness predictions for each filter are compared to a log-scaled COBE/DIRBE data (top panel). The agreement between the Zodiacal feature (S-shaped diffuse feature going from lower left to top right in each map) for filters with the most training data (F606W and F814W) predicted in our models and the COBE/DIRBE data indicates our model can recover the sky background. Filters F435W and F850LP have a smaller training dataset and so do not recover as much of the structure present in the COBE/DIRBE data.}
    \label{fig:allsky}
\end{figure*}

In this section we explore how the results presented above can be used for two applications. First, we discuss how the initial set of space telescope pointing constraints can be used to isolate the observed darkest sky from existing data, and potentially avoid stray light in future missions. Using these initial constraints we isolate observations that are most likely to contain stray light, and observations dominated by Zodiacal Light. We then demonstrate the use of the model to predict the surface brightness over the entire sky and briefly discuss possible applications.
\\
\\
Most space telescopes have hard attitude constraints that dictate how close a target field can be observed with respect to the Earth, Sun and Moon. As described in \autoref{the_start}, for applications where the science target is diffuse light, like the EBL, one strategy is to adopt more stringent attitude constraints to minimise sources of stray light. Below we apply our initial recommended set of telescope attitude constraints to reduce the impact of stray light on HST observations, and to illustrate the potential value of this type of selection.
\\
\\
In \autoref{isolate_stray_light}, we outlined our initial recommended set of telescope attitude constraints to avoid stray light. To explore how these constraints might be used, a Sun Altitude limit of $< -10 ^{\circ}$, a Limb Angle limit of $ > 20^{\circ}$ and a Sun Angle limit $> 80 ^{\circ}$ were applied to the F435W, F606W and F850LP data in the GOODS North field. The resulting subset of sky measurements were averaged and are shown in \autoref{fig:sed_test}. This SED should have very minimal levels of stray light and should only reflect the next most important component in the sky, the Zodiacal Light. When this SED is compared to a scaled Zodiacal Light template, we find good overall agreement in the shape, supporting the idea the data has been isolated from stray light and resembles the predicted Zodiacal Light in this field.
\\
\\
Also in \autoref{fig:sed_test} we show the average sky surface brightness from the subset of data that we expect should have the highest Earthshine stray light contribution based on the results of \autoref{isolate_stray_light}. Observations are selected with a Sun Altitude of $ > 60 ^{\circ}$, a Limb Angle of $ < 40^{\circ}$. These limits are chosen to isolate fields that are most likely to contain stray light, and a larger Limb Angle range is used to ensure data samples are large enough to be statistically significant. A Sun Angle limit $> 80 ^{\circ}$ is also applied to ensure the contribution of Zodiacal Light intensity is approximately consistent between samples of minimal stray light, and maximum stray light. 
\\
\\
The shape of this SED is markedly different than the one with minimal stray light, indicating there is more than just Zodiacal Light in these fields. We found a Zodiacal Light plus Solar template based on the results of \cite{Giavalisco2002} did not fit the SED for wavelengths shorter that $\sim 0.4 \mu$m, so we instead attempted to incorporate an estimate of what a Earthshine spectra might look like. Rayleigh scattered Sunlight is thought to make up a major component of the stray light from Earthshine \citep{Woolf2002}, however we note this is a simplification of an expected Earthshine spectrum which is known to contain many absorption features and other scattering properties from surface features and cloud cover. We fit the data with a Rayleigh scattered modified Solar spectrum (90 degree scattering angle) to mimic Earthshine plus a Zodiacal Light spectrum using a least squares fit of:
\begin{equation}
    F_{\lambda} = \alpha F_{Zodi} + \gamma F_{Earthshine}
\end{equation}
Where $\alpha$, $\gamma$ are simple scaling parameters for each component of the sky SED. As shown in \autoref{fig:sed_test}, this reproduces the same F435W excess seen in images most likely to contain stray light. The depression in F606W may be as a result of absorption features in the Earthshine spectrum in $0.5 - 0.6 \mu$m wavelengths \citep{Woolf2002}. The best fit SED includes an average $30.5 \pm 2.9\%$ Earthshine component across the 3 filters.
\\
\\
The agreement between the template fits and our SEDs of \autoref{fig:sed_test} suggests that the method of applying telescope attitude constraints is indeed effective at reducing or enhancing the impact of stray light. It also suggests that exposures most likely to contain stray light may indeed have observed sky that contains a Rayleigh scattered Earthshine component, possibly originating from scattered Sunlight in the atmosphere, or reflected from surface clouds reaching the telescope aperture. Future work using multiple filters to more finely sampling the sky SED, may be able to identify characteristic traits of the Earth's scattered and reflected SED such as water, oxygen and ozone absorption features \citep{Woolf2002} to confirm the origin of the stray light. 
\\
\\
While our initial set of constraints appear to be promising, we note they were derived using a single field observed with one filter. Moreover, our data are likely not completely free of bad background estimates (see \autoref{data_quality_control}), which will introduce noise into the data. The promising outcome from this work has motivated a more extensive investigation of these issues using essentially the entire HST imaging archive via the SKYSURF Legacy Archival Survey \citep{Windhorst2022}. In this pending work, we have a significantly larger sample size which will allow us to explore numerous other factors (e.g. instrumental contributions to the sky, Galactic dust, etc.) that we have ignored in the present study.
\\
\\
Ultimately the aim of our work is to improve models of the sky to enable low surface brightness science - including observations of the EBL - from LEO. An important result from our work that will support future efforts is the promising use of machine learning to develop a data-driven tool to predict the sky surface brightness for any HST field, which may be extended to future missions.
\\
\\
In \autoref{model_performance} we showed that the empirically-driven model can accurately predict the sky surface brightness of a particular field with a set of telescope attitude parameters to an accuracy of $\approx \beta = 4\%$ for filters with the most training data. Further improvements are likely possible and will be explored in a future work. The model is also shown to predict sky surface brightness estimates for the Zodiacal Light dominated sky to an accuracy better than some existing analytical/physical models.
\\
\\
The benefit of this approach is that a large parameter space can be reduced to one flexible model, and subtle trends not captured in the simple analysis of \autoref{isolate_stray_light} may be teased out across the entire optical/near-IR spectral range. It may also be used to provide improved telescope attitude constraints for LEO space telescopes that can be used to avoid stray light. The method can be extended to future missions with HST-like optics and inform future HST programs.
\\
\\
To further illustrate the potential of this model, we train the model on our entire dataset. This allows us to interpolate data across space and time and fix orbital parameters to see what the entire celestial sphere would look like to HST. This enables us to produce all sky surface brightness maps that represent the sky when contributions from Earthshine, Moonshine and Sunshine are minimised (or maximised) using our derived constraints in \autoref{isolate_stray_light}.
\\
\\
The maps presented in \autoref{fig:allsky} were constructed by fixing the parameters according to our recommended constraints in \autoref{isolate_stray_light} to reduce the impact of stray light. The maps themselves were created by random sampling of the model's celestial sphere to remove any artificial structure that may result from uniformly gridded data. Triangular interpolation is then used to re-sample the results onto identical projections for each filter. The number of predicted points is set corresponding to the sample size of each filter in order to capture the relative uncertainty for each filter. F606W and F814W with the highest sample sizes are composed of 400 predicted points, re-sampled onto a 100 by 100 uniform grid. F435W and F850LP have considerably fewer training data points, and so are re-sampled onto a 50 by 50 grid. Finer grid size and larger sampling numbers do not change the results significantly. 
\\
\\
COBE/DIRBE $1.25\mu$m \footnote{\url{https://lambda.gsfc.nasa.gov/product/cobe/dirbe\_products.cfm}} is included in \autoref{fig:allsky} as a baseline comparison to our HST sky model. A log scale is used to highlight the faint Zodiacal belt in COBE/DRIBE wavelengths. The Zodiacal belt in HST data follows similar trends to DIRBE data. The Zodiacal Light in our model is brighter compared to the Galactic plane, which is marginally detected. This most likely reflects poorer sampling due to the avoidance of crowded fields along the Galactic plane as shown in \autoref{fig:field_locations}. HEALpix maps of these data will be provided on request, and the results can be reproduced following the examples in the {\tt{Skypy}} repository. This comparison illustrates how data-driven models can recover the structure of the sky off the Galactic plane. 
\\
\\
These all-sky maps help illustrate another possible application of this type of model, which is to improve exposure time calculations for future HST observations. This works because we effectively will have a highly-tuned prediction for the sky surface brightness for a given field observed at a particular orbital configuration with HST. If the observations are sky-background limited, more optimal scheduling might take place that are appropriate for exact orbital configuration that has been allocated to a particular program or visit. Future work will be conducted to explore if this will lead to significant efficiency in HST scheduling and perhaps motivate a similar study for future missions.

%% file: 05_conclusion.tex
\section{Conclusions and Future work}

In this study we confirm previous work \cite[e.g.][]{Biretta2003,Giavalisco2002,Shaw1998} that stray light contamination from Earthshine is an important factor that influences the intensity of sky in HST ACS filters. We find that in worst case scenarios (i.e., when HST is pointing relatively close to the daytime Earth's limb), the measured sky can increase by as much as $4 \times$ the expected Zodiacal Light surface brightness. Furthermore it varies by as much as $ \sim 30\%$ between exposures due to the changes in the telescope attitude with respect to sources of stray light. Stray light from Moonshine is still likely to be a problem but we do not see strong evidence in the limited data set we analysed. Here we do not attempt to decouple stray light from direct Sunlight from Zodiacal Light. The extent and variability of Earthshine stray light in HST ACS images as illustrated by this work underscore the need for space telescopes that are designed and specialised to reduce the significant impact of stray light for low surface brightness and EBL observations from LEO.
\\
\\
Fortunately, as in \cite{Biretta2003} we find that there are certain restrictions on orbital parameters that significantly minimise stray light contributions to the observed sky. We find that Sun Altitude and Limb Angle are most important factors in determining the potential for stray light contamination in HST images, and should both be constrained to $< -10^{\circ}$ and $> 20^{\circ}$ respectively in order to minimise the contribution of stray light to HST observations. We also present evidence that satellite weather data may be useful in identifying observing conditions that are most likely to result in high Earthshine levels, however we note that more data is needed to confirm these findings. 
\\
\\
We also present a new XGBoost machine learning model trained on this test data set, and demonstrate its ability to accurately predict the sky surface brightness over a range of different observing conditions. We demonstrate that this empirical model is able to achieve an overall accuracy that is better than or comparable to existing sky Zodiacal Light models. For the dataset with the largest training sample F814W, we are able to recover the total night time sky surface brightness to an accuracy of $\sim 4\%$, which is comparable to the fraction percentage of the lower limit of the EBL from \cite{Driver2016}. 
\\
\\
We conclude with a note that we will extend this study to a much larger dataset. We will utilise the full HST legacy archive via the SKYSURF Legacy Archival Survey \citep{Windhorst2022, Carleton2022} for WFC3 and a more extensive ACS data set. This will enable us to extend our work to a more thoroughly sampled parameter space and work towards decoupling sources of stray light from the spatial dependencies of the Zodiacal Light. This will enable us to probe the origins and structure of the Zodiacal Cloud, and improve Zodiacal Light models for future mission such as JWST, SphereX, Euclid and Roman. In this work we confirm that stray light contamination from Earthshine impacts the intensity of the sky from LEO, and as a result this excess diffuse light may help to explain some of the observational discrepancies between direct EBL measurements from LEO and estimates derived by galaxies counts. Ultimately, furthering this work with SKYSURF will aid in making progress towards secure measurements of the EBL.